\documentclass[twocolumn]{article}
\usepackage[utf8]{inputenc}
\usepackage{cite}
\usepackage{authblk}
\usepackage{mathtools, cuted}
\usepackage{graphicx}
\usepackage{cancel}
\usepackage{xcolor}
\usepackage{longtable}
\usepackage{hyperref}
\usepackage{widetext}

\def\Eq#1{Eq.~(\ref{#1})}
\def\beq{\begin{equation}}
\def\eeq{\end{equation}}
\def\bea{\begin{eqnarray}}
\def\eea{\end{eqnarray}}
\def\nn{\nonumber}

\def\ket#1{|{#1}\rangle}
\def\bra#1{\langle{#1}|}
\def\id{\boldsymbol I}

\title{\textbf{Four-loop scattering amplitudes through \\the loop-tree duality}}
\author{Selomit Ram\'irez-Uribe \footnote{norma.selomit.ramirez@ific.uv.es} }
\affil{Instituto de F\'isica Corpuscular, Universitat de Val\`encia -- Consejo Superior de Investigaciones Cient\'ificas, 
Parc Cient\'ific, E-46980 Paterna, Valencia, Spain.}
\affil{Facultad de Ciencias de la Tierra y el Espacio,
Universidad Aut\'onoma de Sinaloa, \\Ciudad Universitaria, CP 80000 Culiac\'an, Mexico.}
\affil{Facultad de Ciencias F\'isico-Matem\'aticas,
Universidad Aut\'onoma de Sinaloa, \\
Ciudad Universitaria, CP 80000 Culiac\'an, Mexico.}

\date{}
\begin{document}
\maketitle
\null \vspace{-15cm}

\begin{strip}
A general outlook is presented on the study of multiloop topologies appearing for the first time at four loops. A unified description and representation of this family is provided, the so-called N$^4$MLT universal topology. Based on the Loop-Tree Duality framework, we discuss the dual opening of this family and expose the relevance of a causal representation.  
We explore an alternative procedure for the search of causal singular configurations of selected N$^4$MLT Feynman diagrams through the application of a modified Grover's quantum algorithm.
\end{strip}

\section{Introduction}
Precision modelling in particle physics is strongly supported by perturbative Quantum Field Theory, hence the importance of progressing towards higher perturbative orders. One of the most relevant difficulties in the topic is the description of quantum fluctuations at high-energy scattering processes. Taking into account that computation of multiloop scattering amplitudes requires an appropriate treatment of loop diagrams, we based the multiloop topology analysis on the Loop-Tree Duality (LTD) formalism~\cite{Catani:2008xa,Rodrigo:2008fp,Bierenbaum:2010cy,Bierenbaum:2012th,Tomboulis:2017rvd,Runkel:2019yrs,Capatti:2019ypt}.

The LTD is an innovative technique that opens any loop diagram into a sum of connected trees having as main property the integrand-level distinction between physical and nonphysical singularities Ref.~\cite{Buchta:2014dfa,Aguilera-Verdugo:2019kbz}. Since the introduction of this formalism, a great effort has been dedicated to understand it in depth, and many interesting features have been found~\cite{Buchta:2015wna,Buchta:2015xda,Driencourt-Mangin:2019yhu,Capatti:2019edf,Jurado:2017xut,Beneke:1997zp,Driencourt-Mangin:2017gop,Plenter:2019jyj,Plenter:2020lop}. Two important applications were addressed through the LTD: local renormalization strategies~\cite{Driencourt-Mangin:2019aix,Prisco:2020kyb} and the cross-section computation in four space-time dimensions at integrand level through the Four Dimensional Unsubtraction~\cite{Hernandez-Pinto:2015ysa,Sborlini:2016gbr,Sborlini:2016hat,Driencourt-Mangin:2019sfl}

Recently, a significant development was presented, a manifestly causal LTD reformulation to all orders~\cite{Verdugo:2020kzh}. The strategy adopted was based on the application of nested residues~\cite{Aguilera-Verdugo:2020fsn} which allows to find more compact and manifestly causal dual expressions. The analysis considered a selected set of multiloop topologies, those appearing for the first time at one loop, two loops (MLT) and three loops (NMLT and N$^2$MLT). Since then, LTD has undergone a remarkable and exciting evolution~\cite{snowmass2020,Aguilera-Verdugo:2020kzc,Sborlini:2021owe,Bobadilla:2021pvr,TorresBobadilla:2021ivx,Ramirez-Uribe:2020hes,Ramirez-Uribe:2021ubp}. 

This paper presents an overview of the study of multiloop topologies that first appear at four loops (N$^4$MLT)~\cite{Ramirez-Uribe:2020hes}, including a unified representation, the dual opening and the causal representation. Related to the causal LTD representation, we explore the application of a modified Grover's quantum algorithm~\cite{Ramirez-Uribe:2021ubp} to solve the problems associated to the identification of singular causal configurations of the N$^4$MLT topologies.

\section{Loop-Tree Duality}
A general scattering amplitude with $P$ external legs is written in accordance with Ref.~\cite{Ramirez-Uribe:2020hes,Ramirez-Uribe:2021ubp} as
\begin{align} \label{eq:LamplitudeN}
\mathcal{A}_F^{(L)}= \int_{\ell_1, \ldots, \ell_L}  \mathcal{N}( \{ \ell_s\}_L,  \{ p_j\}_P) \, \prod_{i=0}^{n}G_F(q_i)\, ,
\end{align}
where the integration measure in dimensional regularization~\cite{Bollini:1972ui,tHooft:1972tcz} is denoted as $\int_{\ell_s}=-\imath\mu^{4-d}\int \frac{d^d\ell_s}{(2\pi)^d}$ with $d$ the number of space-time dimensions.  

At one loop, the LTD representation is obtained applying the Cauchy's residue theorem to \Eq{eq:LamplitudeN}, i.e., integrating over one component of the $L$ loop momenta. Having a multiloop scattering amplitude scenario, the LTD representation is calculated based on the evaluation of nested residues ~\cite{Verdugo:2020kzh,Aguilera-Verdugo:2020fsn}, 
\begin{align}
\label{eq:nested}
&\mathcal{A}_D^{(L)}(1,\ldots, r; r+1,\dots, n)=-2\pi \imath \times \\
&\sum_{i_r \in r} {\rm Res} (\mathcal{A}_D^{(L)}(1, \ldots, r-1;r, \ldots, n), {\rm Im}(\eta\cdot q_{i_r})<0)\, , \nn
\end{align}
where the arguments to the left of the semicolon represent the sets containing an on-shell propagator and the ones located to the right are those with all the propagators off shell.
The futurelike vector $\eta$ indicates the loop momenta component to be integrated, in our case the selection is $\eta^{\mu}=(1,{\bf 0})$. Integrating over the energy component gives the advantage of working in the Euclidean integration domain of the loop three-momenta space instead of a Minkowky space. 

\section{N$^4$MLT topology}
\begin{figure}[h!]
    \centering
    \includegraphics[scale=0.5]{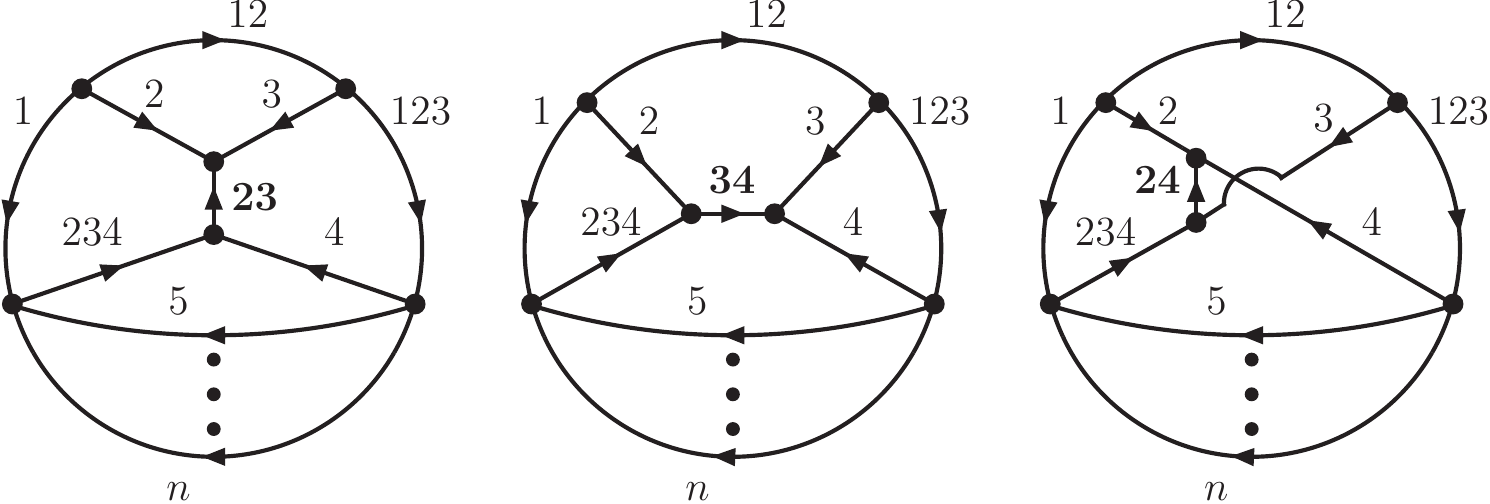}
    \caption{Representative diagrams of the N$^4$MLT family. From left to right the diagrams correspond to the $t$, $s$ and $u$ channels.}
    \label{fig:family}
\end{figure}

The loop topologies appearing for the first time at four loops correspond to the N$^3$MLT and N$^4$MLT topologies, representing those with $L+4$ and $L+5$ sets of propagators respectively.
The N$^4$MLT family naturally includes the N$^3$MLT, and it can be fully studied through three main topologies depicted in Figure~\ref{fig:family}.

\subsection{Universal topology}
The topologies shown in Figure~\ref{fig:family} are interpreted as the $t$-, $s$- and $u$-kinematic channels, enabling to provide a unified description. 
Given the similarities among the three main topologies and with the purpose to obtain a general expression, a current $J$ is defined as $J \equiv 23\cup 34\cup 24$. This statement allows to merge the three representative topologies into a single one, the N$^4$MLT {\it universal topology} written as
\begin{align}\label{eq:universal}
&{\cal A}^{(L)}_{\rm N^4MLT}={\cal A}^{(L)}_{F} (1, \ldots, L+1, 12, 123, 234, J)~.  
\end{align}

\begin{figure}[h!]
    \centering
    \includegraphics[scale=0.5]{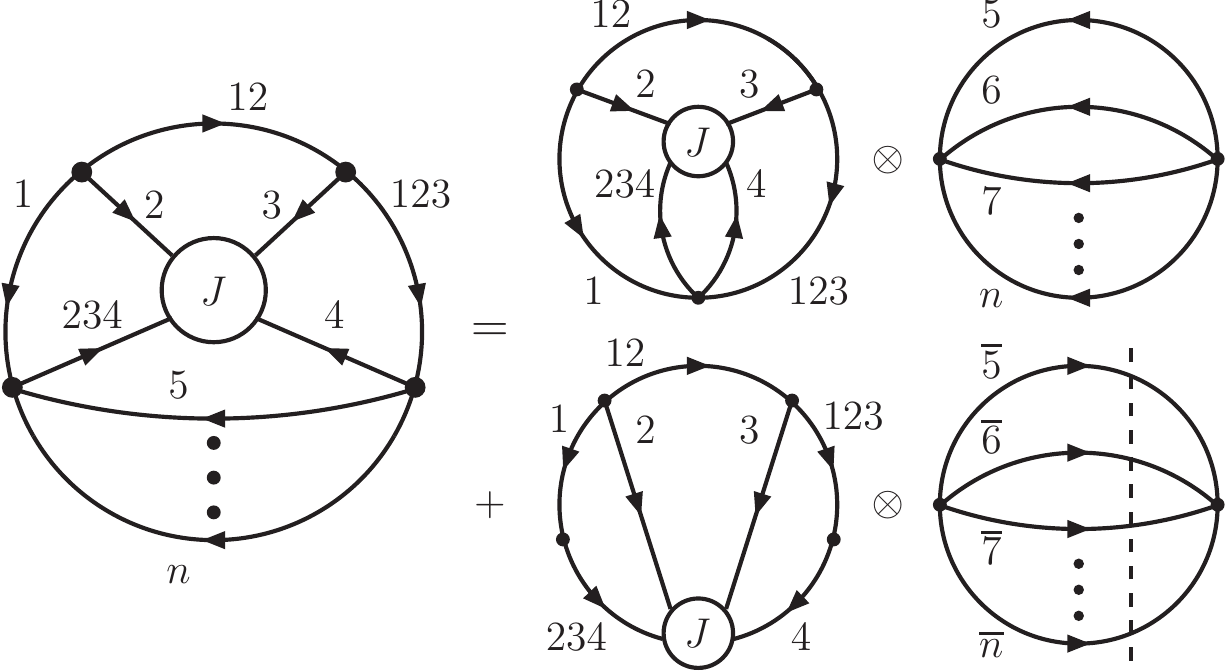}
    \caption{Illustrative representation of the N$^4$MLT universal topology dual opening. Only the cut on the last MLT subtopology is indicated.}
    \label{fig:master}
\end{figure}

The nested residues evaluation of \Eq{eq:universal} gives the dual opening depicted in Figure~\ref{fig:master}, and stands as
\begin{align}
\label{eq:master}
&{\cal A}^{(L)}_{\rm N^4MLT} (1, \ldots, L+1, 12, 123, 234, J)  \\
& =  {\cal A}^{(4)}_{\rm N^4MLT} (1, 2, 3, 4, 12, 123, 234, J) \otimes {\cal A}_{\rm MLT}^{(L-4)} (5, \dots, n) \nn \\
& + {\cal A}^{(3)}_{\rm N^2MLT} (1\cup 234, 2, 3, 4\cup 123, 12, J) \otimes {\cal A}_{\rm MLT}^{(L-3)} (\overline 5, \dots, \overline{n} )~, \nn 
\end{align}
where the bar indicates the change on the original momentum flow. 
The computation of ${\cal A}_{\rm MLT}^{(L-4)} (5, \dots, n)$ and ${\cal A}_{\rm MLT}^{(L-3)} (\overline 5, \dots, \overline{n} )$ in \Eq{eq:master} is according to Ref.~\cite{Verdugo:2020kzh}. The terms ${\cal A}^{(4)}_{\rm N^4MLT}$ and ${\cal A}^{(3)}_{\rm N^2MLT}$ are opened following a factorization identity described in terms of known subtopologies; this takes into account all feasible arrangements with four and three on-shell conditions, respectively.

The LTD representation of the N$^4$MLT topologies are obtained through the factorized dual expression shown in \Eq{eq:master}. The analytical expressions are provided in Ref.~\cite{Ramirez-Uribe:2020hes}, as it was expected, they satisfy the condition of the absence of disconnected trees.

\subsection{Causal representation}
The fundamental distinction between the direct and causal LTD representations is the presence or absence of noncausal singularities.
The interest in this topic is motivated by the advantage of working with a causal representation. 
To exhibit the impact of it, an analysis of a N$^3$MLT diagram is presented.

\begin{figure}[h!]
    \centering
    \includegraphics[scale=0.39]{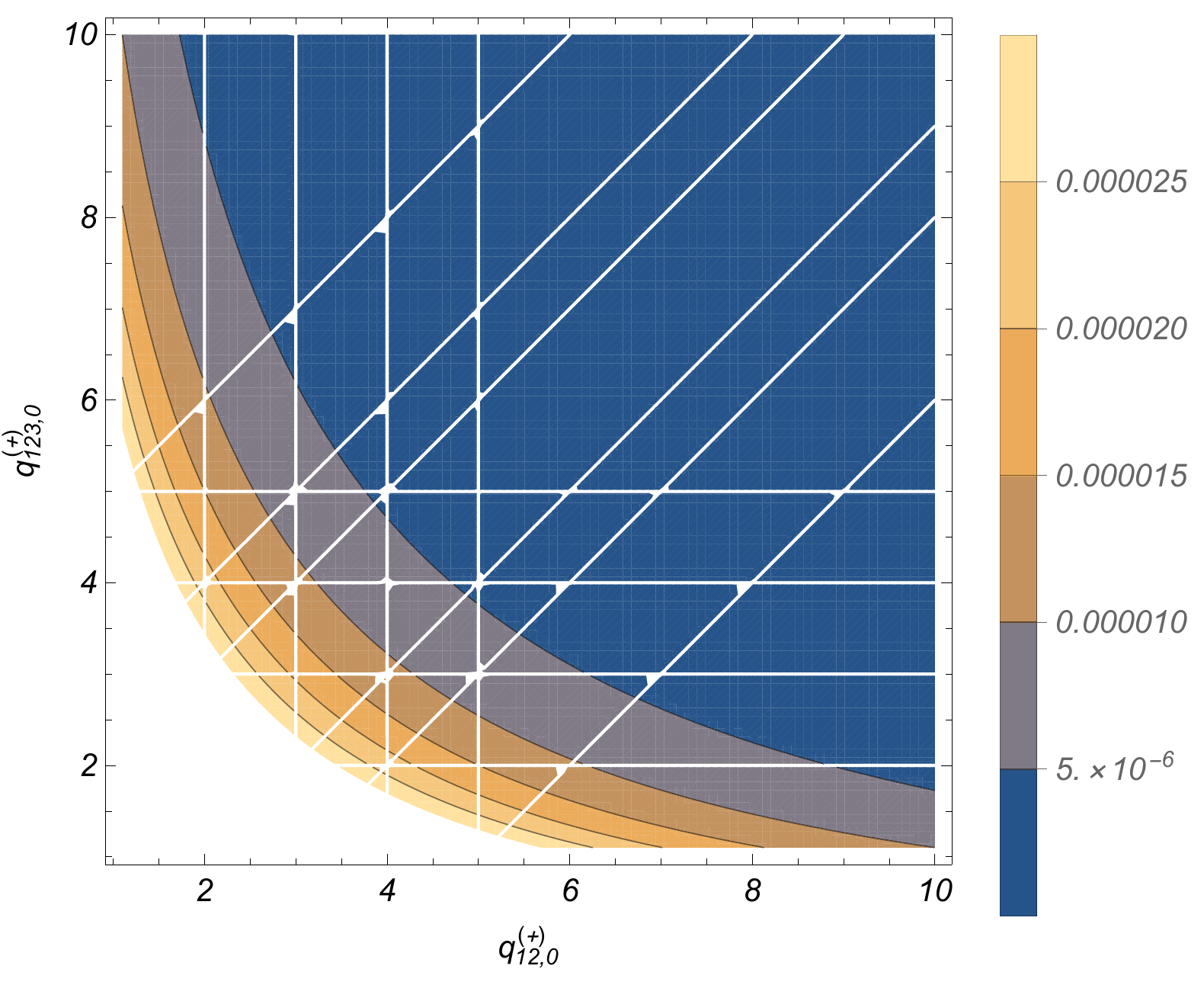}
    \caption{Singular behaviour at integrand level of the direct LTD representation of the N$^3$MLT vacuum diagram.}
    \label{fig:threshold}
\end{figure}

In Figure~\ref{fig:threshold} is shown the integrand of the direct LTD representation of a N$^3$MLT diagram as a function of $q_{12,0}^{(+)}$ and $q_{123,0}^{(+)}$. We can observe the white lines standing for the location of noncausal singularities.

For a better understanding of the noncausal thresholds in Figure~\ref{fig:threshold}, one of the singularities is analized through a comparative between the direct and causal representations. To obtain the causal representation of the N$^3$MLT vacuum diagram the procedure exposed in Ref.~\cite{Aguilera-Verdugo:2020kzc} is followed, having as a key strategy the search of causal compatibility among four entangled thresholds.

\begin{figure}[h!]
    \centering
    \includegraphics[width=110px]{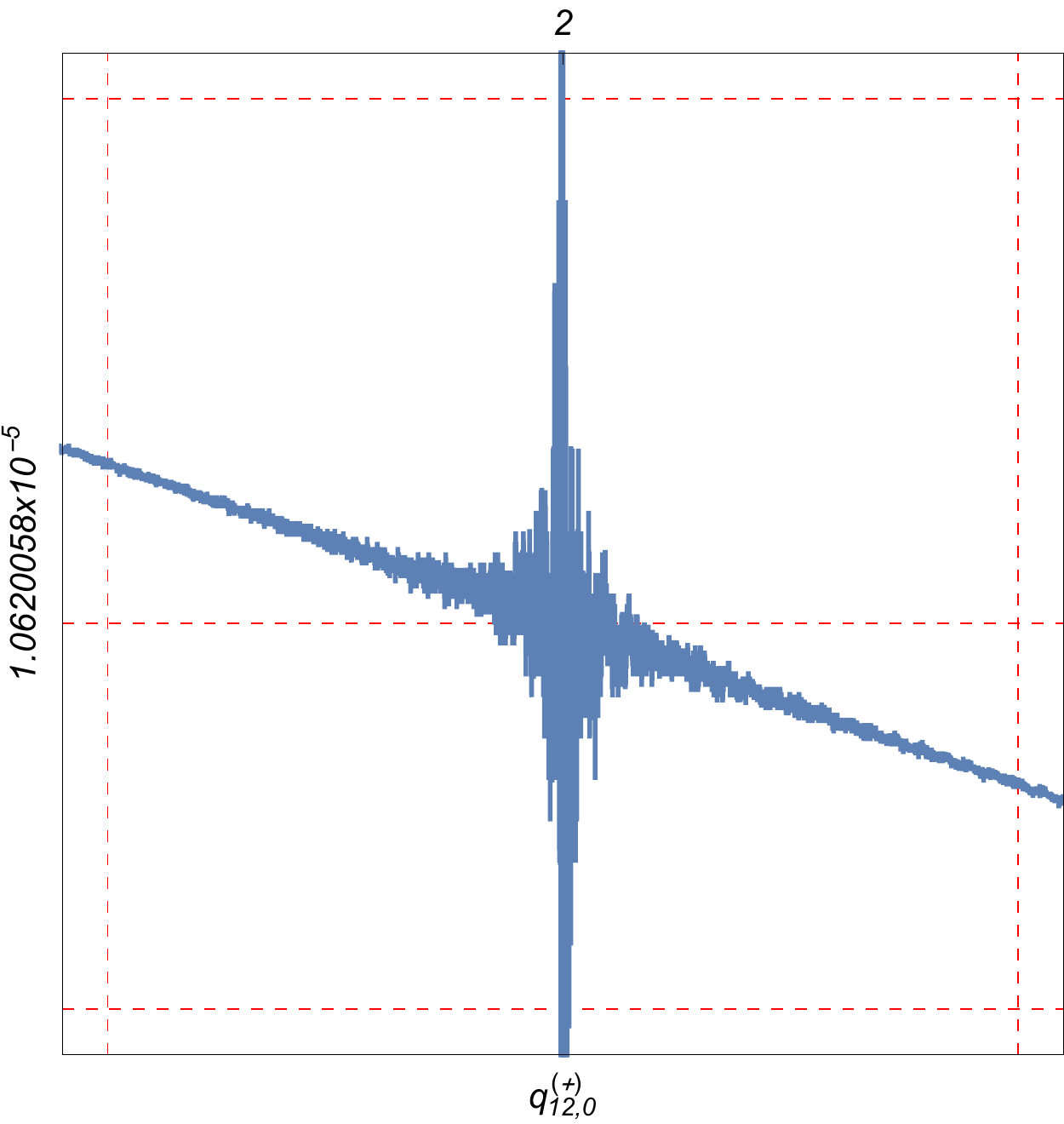}
    \includegraphics[width=110px]{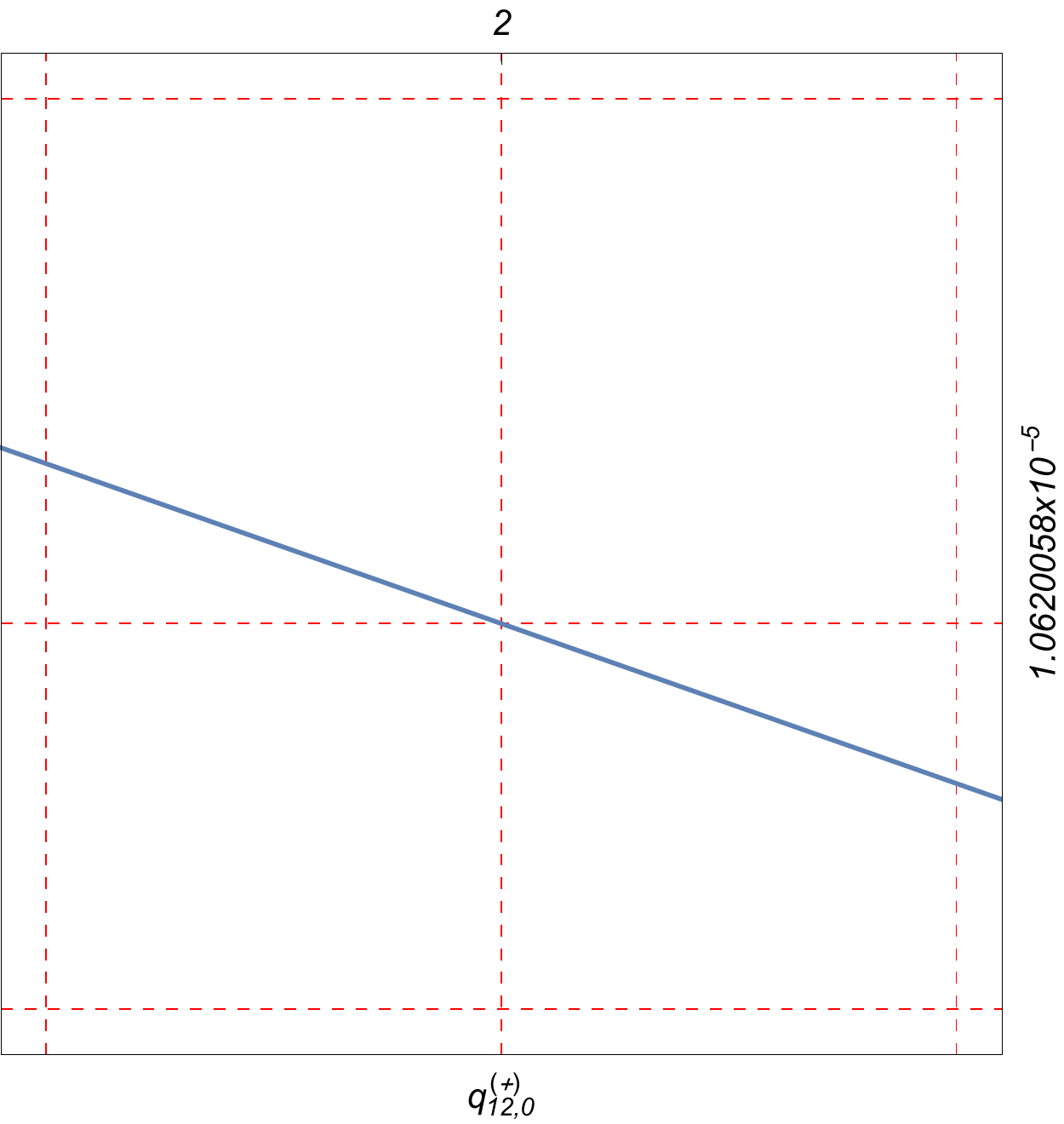}
    \caption{Numerical performance of direct (left) and causal (right) N$^3$MLT integrand.}
    \label{fig:contrast}
\end{figure}

The analysis of the singularity arising from the N$^3$MLT vacuum diagram is realized by taking the on-shell energy $q_{123,0}^{(+)}$ as fixed and scanning over $q_{12,0}^{(+)}$; the evaluations of direct and causal N$^3$MLT integrands are illustrated in Figure~\ref{fig:contrast}. As we can notice, the right plot shows a desirable numerical stability; the left plot exhibits numerical instabilities coming from noncausal singularities.

Going forward to the causal representations of the three main topologies of the N$^4$MLT family, the same procedure is followed with the particularity that causal compatibility has to be fulfill among five thresholds. All the explicit results and details about the causal analysis are fully presented in Ref.~\cite{Ramirez-Uribe:2020hes}. 

\section{Feynman integrals through a quantum algorithm}
\begin{figure}[h!]
    \centering
    \includegraphics[scale=0.27]{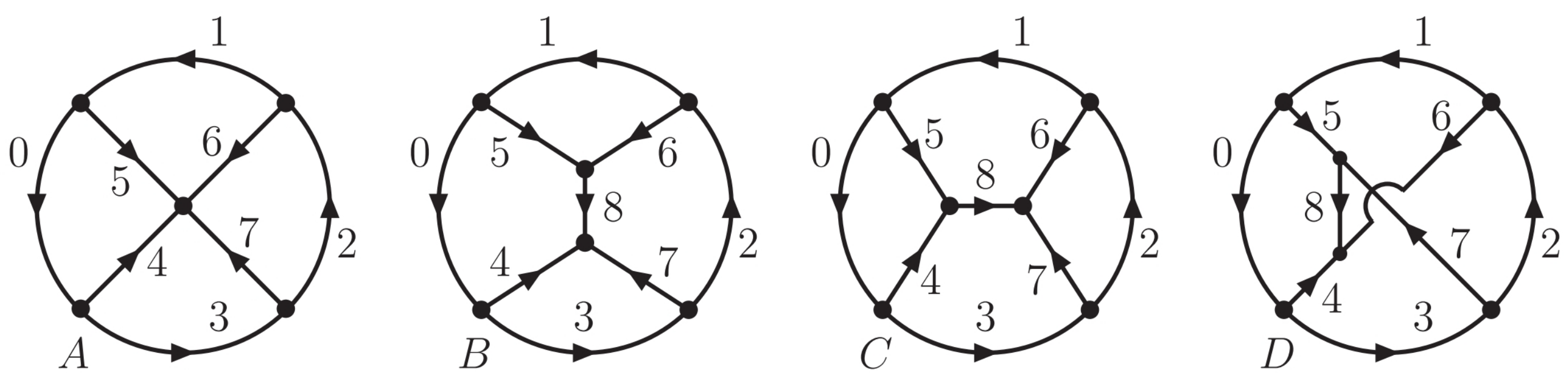}
    \caption{Selected four-loop topologies. The arrows' direction stand for the initial flow configuration of the diagrams.}
    \label{fig:Diagram_QC}
\end{figure}

A natural association between Feynman loops integrals and quantum computing is based on the fact that a Feynman propagator can be represented in terms of a qubit. A propagator has only two possible on-shell states, $\ket{1}$ representing those states with the initial flow configuration and $\ket{0}$ for those with inverse flow orientation. The specific four-loop diagrams analyzed and the initial configuration are shown in Figure~\ref{fig:Diagram_QC}.  

An important problem to solve in the causal representation context is the  identification of causal singular configurations. An alternative to deal with this difficulty is to understand it from a quantum computing point of view, as a query over unstructured datasets~\cite{Boyer:1996zf}. The scheme explored was the application of Grover's quantum algorithm~\cite{Grover:1997fa}.

Grover's quantum algorithm relies in three main ingredients: uniform superposition, oracle operator and diffusion operator.

\begin{enumerate}

\item The uniform superposition of the $N=2^n$ possible states is given by
\beq\label{eq:superposition}
\ket{q}= \frac{1}{\sqrt{N}} \sum_{x=0}^{N-1} \ket{x}~, 
\eeq
where $n$ denotes the number of Feynman propagators. \Eq{eq:superposition} can also be written as the superposition of the winning ($\ket{w}$) and orthogonal ($\ket{q_\perp}$) states, 
\beq
\ket{q} =  \cos \theta \, \ket{q_\perp} + \sin\theta \, \ket{w}~.
\eeq
The mixing angle between those states is given by $\theta = \arcsin \sqrt{r/N}$, where $r$ is the number of elements of the winning state.

\item The oracle operator, $U_w = \id - 2\ket{w} \bra{w}$, flips the state $\ket{x}$ if $x\in w$ and leaves it unchanged otherwise.

\item The diffusion operator, $U_q = 2 \ket{q} \bra{q} - \id$, performs a reflection around the initial state $\ket{q}$ in order to amplify the probability of the winning states. 
\end{enumerate}
An iterated application of ii) and iii) $t$ times leads to
$(U_q U_w )^t \ket{q} = \cos \theta_t \, \ket{q_\perp } +  \sin \theta_t \, \ket{w}$,
with $\theta_t = (2t +1) \, \theta$.

To define a proper number of iterations, the mixing angle plays a crucial role, i.e., the proportion between the number of elements in the winning states and the total states. If $r \leq N/4$, the standard Grover's algorithm is a promising approach, on the contrary, its amplitude amplification performance is not satisfactory.

In the case of the N$^4$MLT family, we know for classical computation that the number of causal singular configurations is near half of the total of possible states~\cite{Ramirez-Uribe:2020hes}. A clever modification to reduce the number of solutions proposed in Ref.~\cite{Ramirez-Uribe:2021ubp} is based on the fact that given one causal solution, the mirror configuration resulting from the momentum flow reversal is also a causal solution.
This modification is achieved by fixing one of the qubits, reducing the number of solutions by half.

The proposed quantum algorithm requires three registers with one additional qubit needed as a marker in the Grover's oracle. The register $q_i$ stands for the state of the $n$ propagators. The second register, $\ket{c}$, contains the binary clauses label as $c_{ij}$ or $\bar{c}_{ij}$. These binary clauses allows to prove if the flow orientation of two adjacent propagators are in the same direction and are defined as
\beq 
c_{ij} \equiv (q_i = q_j),\quad \bar c_{ij} \equiv (q_i \ne q_j)~,
\eeq
with $i,j \in\{0, \ldots, n-1\}$.
Finally, the $\ket{a}$ register stores loop clauses. These are used to validate if all subloop configurations form a cyclic circuit applying a multi-Toffoli gate comparing qubits from $\ket{c}$.

The general structure of the algorithm is described below:
\begin{itemize}
\item The starting point is to initialize the registers described above.  First of all, the uniform superposition is applied to the qubits encoding the propagators through the Hadamard gate, $\ket{q} = H^{\otimes n} \ket{0}$. The registers $\ket{a}$, $\ket{c}$ are set to $\ket{0}$ and the qubit associated to the Grover's marker is set to the Bell state, $\ket{out_0} = \left(\ket{0} - \ket{1}\right)/\sqrt{2}$.

\item To compare two adjacent lines, $\bar c_{ij}$ needs two CNOT gates which operate between $q_i$, $q_j$ and a qubit in the $\ket{c}$ register. If the binary clause to be implement is $c_{ij}$, an extra XNOT gate is required to perform on the corresponding qubit in $\ket{c}$. 

\item The oracle operator requires a function, $f(a,q)$, defined in such a way that if the winning state conditions are satisfied then $f(a,q) = 1$, if not $f(a,q) = 0$. In addition to the causal restrictions, this function considers an arbitrary qubit as fixed. Once this function has been set, the oracle is applied as 
\bea
U_w \ket{q} \ket{c} \ket{a} \ket{out_0} = \ket{q} \ket{c} \ket{a} \ket{out_0 \otimes f(a,q)}
\eea
with
\beq
\ket{out_0 \otimes f(a,q)}=\left\lbrace
	\begin{aligned}
	-\ket{out_0},& \quad {\rm if}~ q \in w  \\
	\ket{out_0},& \quad {\rm if}~ q \not\in w 
	\end{aligned}
\right.
\eeq
After marking the causal states the operations of the oracle are implemented in reverse order.

\item Before measuring, the diffuser operator is applied to $\ket{q}$. This operator is taken from the documentation provided in the IBM Qiskit website (\texttt{https://qiskit.org/}). 
\end{itemize}

The adapted Grover's quantum algorithm was applied to the multiloop N$^3$MLT, $t$, $s$ and $u$ channels shown in Figure~\ref{fig:Diagram_QC}. The implementation was performed on the IBM's quantum simulator provided by Qiskit framework (upper limit 32 qubits). For the $u$ channel the algorithm required 33 qubits, more than Qiskit capacity. In this case the algorithm was implemented within QUTE Testbed framework~\cite{alonso_raul_2021_5561050}.

\begin{figure}[h!]
    \centering
    \includegraphics[scale=0.13]{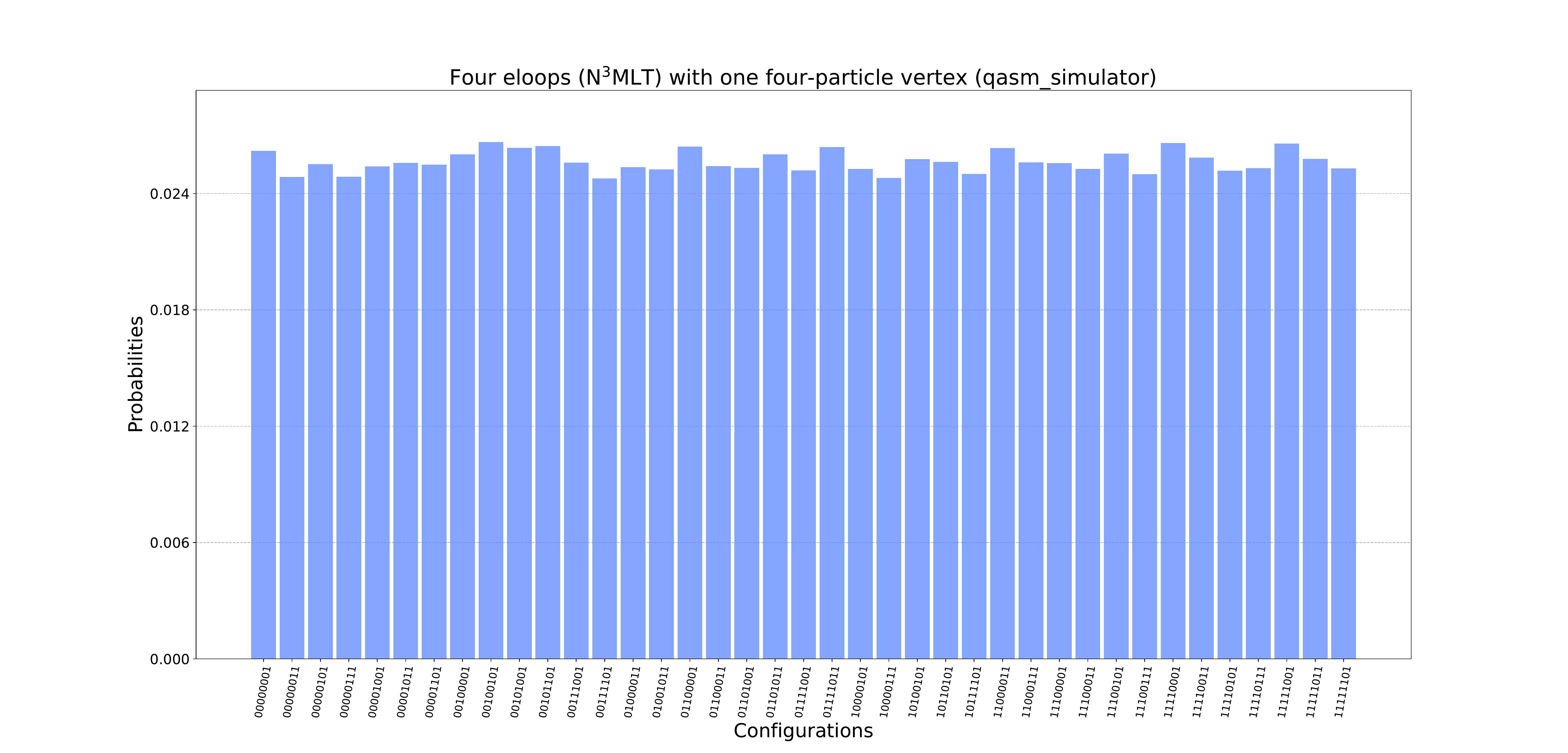}
    \caption{Causal states probabilities for N$^3$MLT.}
    \label{fig:Prob}
\end{figure}

The proposed algorithm successfully identified all the causal singular configurations. To expose the performance of the algorithm, in Figure~\ref{fig:Prob} is shown the output of the probabilities of the causal singular states associated to the N$^3$MLT diagram.

\section{Conclusions}
There has been analyzed an extension in complexity of the LTD reformulation enabling to obtain a general expression to describe any scattering amplitude up to four loops, the N$^4$MLT universal topology. 

The dual opening of the universal topology is expressed in a very compact way as a factorization of simpler subtopologies. It is known that the direct LTD representation can be rewritten in terms of only causal propagators, furthermore, interpreting them in terms of entangled thresholds is the key to extract the causal LTD representation. It was emphasized the importance of the causal representation, showing a desirable efficiency over numerical evaluation of multiloop scattering amplitudes.

A modified Grover's quantum algorithm has been applied to the N$^3$MLT and N$^4$MLT topologies, obtaining the causal singular configurations of the selected multiloop Feynman integrals. The application of quantum algorithms related to the LTD formalism is a promising tool that we intend to explore further in order to improve the computation of multiloop scattering amplitudes.

\section*{Acknowledgments}
I am very grateful with R. Hern\'andez-Pinto, G. Rodrigo and G. Sborlini for all the support and guidance through the development of this work. I would like to thank CTIC for granting me access to their simulator Quantum Testbed (QUTE) and IBMQ.
Support for this work has been received in part by MCIN/AEI/10.13039/501100011033, Grant No. PID2020-114473GB-I00, COST Action CA16201 PARTICLEFACE, Project No. A1- S-33202 (Ciencia B\'asica), Consejo Nacional de Ciencia y Tecnología and Universidad Aut\'onoma de Sinaloa.


\begin{thebibliography}{9}

\bibitem{Catani:2008xa}
S.~Catani, T.~Gleisberg, F.~Krauss, G.~Rodrigo and J.-C. Winter, \emph{{From
  loops to trees by-passing Feynman's theorem}},
  \href{http://dx.doi.org/10.1088/1126-6708/2008/09/065}{\emph{JHEP} {\bf 09}
  (2008) 065}, [\href{http://arxiv.org/abs/0804.3170}{{\tt 0804.3170}}].
  
  
\bibitem{Rodrigo:2008fp}
G.~Rodrigo, S.~Catani, T.~Gleisberg, F.~Krauss and J.~C.~Winter, \emph{{From multileg loops to trees (by-passing Feynman's Tree Theorem)}},
\href{https://doi.org/10.1016/j.nuclphysbps.2008.09.114}{\emph{Nucl. Phys. B Proc. Suppl.}{\bf 183} (2008) 262-267},
[\href{http://arxiv.org/abs/0807.0531}{{\tt 0807.0531}}].

\bibitem{Bierenbaum:2010cy}
I.~Bierenbaum, S.~Catani, P.~Draggiotis and G.~Rodrigo, \emph{{A Tree-Loop
  Duality Relation at Two Loops and Beyond}},
  \href{http://dx.doi.org/10.1007/JHEP10(2010)073}{\emph{JHEP} {\bf 10} (2010)
  073}, [\href{http://arxiv.org/abs/1007.0194}{{\tt 1007.0194}}].

\bibitem{Bierenbaum:2012th}
I.~Bierenbaum, S.~Buchta, P.~Draggiotis, I.~Malamos and G.~Rodrigo,
  \emph{{Tree-Loop Duality Relation beyond simple poles}},
  \href{http://dx.doi.org/10.1007/JHEP03(2013)025}{\emph{JHEP} {\bf 03} (2013)
  025}, [\href{http://arxiv.org/abs/1211.5048}{{\tt 1211.5048}}].

\bibitem{Tomboulis:2017rvd}
E.~Tomboulis, \emph{{Causality and Unitarity via the Tree-Loop Duality
  Relation}}, \href{http://dx.doi.org/10.1007/JHEP05(2017)148}{\emph{JHEP} {\bf
  05} (2017) 148}, [\href{http://arxiv.org/abs/1701.07052}{{\tt 1701.07052}}].

\bibitem{Runkel:2019yrs}
R.~Runkel, Z.~Sz\H{o}r, J.~P. Vesga and S.~Weinzierl, \emph{{Causality and
  loop-tree duality at higher loops}},
  \href{http://dx.doi.org/10.1103/PhysRevLett.122.111603,
  10.1103/PhysRevLett.123.059902}{\emph{Phys. Rev. Lett.} {\bf 122} (2019)
  111603}, [\href{http://arxiv.org/abs/1902.02135}{{\tt 1902.02135}}].

\bibitem{Capatti:2019ypt}
Z.~Capatti, V.~Hirschi, D.~Kermanschah and B.~Ruijl, \emph{{Loop-Tree Duality
  for Multiloop Numerical Integration}},
  \href{http://dx.doi.org/10.1103/PhysRevLett.123.151602}{\emph{Phys. Rev.
  Lett.} {\bf 123} (2019) 151602}, [\href{http://arxiv.org/abs/1906.06138}{{\tt
  1906.06138}}].

\bibitem{Buchta:2014dfa}
S.~Buchta, G.~Chachamis, P.~Draggiotis, I.~Malamos and G.~Rodrigo, \emph{{On
  the singular behaviour of scattering amplitudes in quantum field theory}},
  \href{http://dx.doi.org/10.1007/JHEP11(2014)014}{\emph{JHEP} {\bf 11} (2014)
  014}, [\href{http://arxiv.org/abs/1405.7850}{{\tt 1405.7850}}].

\bibitem{Aguilera-Verdugo:2019kbz}
J.~J. Aguilera-Verdugo, F.~Driencourt-Mangin, J.~Plenter,
  S.~Ram{\'\i}rez-Uribe, G.~Rodrigo, G.~F. Sborlini et~al., \emph{{Causality,
  unitarity thresholds, anomalous thresholds and infrared singularities from
  the loop-tree duality at higher orders}},
  \href{http://dx.doi.org/10.1007/JHEP12(2019)163}{\emph{JHEP} {\bf 12} (2019)
  163}, [\href{http://arxiv.org/abs/1904.08389}{{\tt 1904.08389}}].

\bibitem{Buchta:2015xda}
S.~Buchta, \emph{{Theoretical foundations and applications of the Loop-Tree
  Duality in Quantum Field Theories}}.
\newblock PhD thesis, Valencia U., 2015.
\newblock \href{http://arxiv.org/abs/1509.07167}{{\tt 1509.07167}}.

\bibitem{Buchta:2015wna}
S.~Buchta, G.~Chachamis, P.~Draggiotis and G.~Rodrigo, \emph{{Numerical
  implementation of the loop--tree duality method}},
  \href{http://dx.doi.org/10.1140/epjc/s10052-017-4833-6}{\emph{Eur. Phys. J.}
  {\bf C77} (2017) 274}, [\href{http://arxiv.org/abs/1510.00187}{{\tt
  1510.00187}}].

\bibitem{Driencourt-Mangin:2019yhu}
F.~Driencourt-Mangin, G.~Rodrigo, G.~F. Sborlini and W.~J. Torres~Bobadilla,
  \emph{{On the interplay between the loop-tree duality and helicity
  amplitudes}},  \href{http://arxiv.org/abs/1911.11125}{{\tt 1911.11125}}.

\bibitem{Capatti:2019edf}
Z.~Capatti, V.~Hirschi, D.~Kermanschah, A.~Pelloni and B.~Ruijl,
  \emph{{Numerical Loop-Tree Duality: contour deformation and subtraction}},
  \href{http://dx.doi.org/10.1007/JHEP04(2020)096}{\emph{JHEP} {\bf 04} (2020)
  096}, [\href{http://arxiv.org/abs/1912.09291}{{\tt 1912.09291}}].


\bibitem{Jurado:2017xut}
J.~L. Jurado, G.~Rodrigo and W.~J. Torres~Bobadilla, \emph{{From Jacobi
  off-shell currents to integral relations}},
  \href{http://dx.doi.org/10.1007/JHEP12(2017)122}{\emph{JHEP} {\bf 12} (2017)
  122}, [\href{http://arxiv.org/abs/1710.11010}{{\tt 1710.11010}}].

\bibitem{Beneke:1997zp}
M.~Beneke and V.~A. Smirnov, \emph{{Asymptotic expansion of Feynman integrals
  near threshold}},
  \href{http://dx.doi.org/10.1016/S0550-3213(98)00138-2}{\emph{Nucl. Phys.}
  {\bf B522} (1998) 321--344}, [\href{http://arxiv.org/abs/hep-ph/9711391}{{\tt
  hep-ph/9711391}}].

\bibitem{Driencourt-Mangin:2017gop}
F.~Driencourt-Mangin, G.~Rodrigo and G.~F. Sborlini, \emph{{Universal dual
  amplitudes and asymptotic expansions for $gg\rightarrow H$ and $H\rightarrow
  \gamma \gamma $ in four dimensions}},
  \href{http://dx.doi.org/10.1140/epjc/s10052-018-5692-5}{\emph{Eur. Phys. J.
  C} {\bf 78} (2018) 231}, [\href{http://arxiv.org/abs/1702.07581}{{\tt
  1702.07581}}].

\bibitem{Plenter:2019jyj}
J.~Plenter, \emph{{Asymptotic Expansions Through the Loop-Tree Duality}},
  \href{http://dx.doi.org/10.5506/APhysPolB.50.1983}{\emph{Acta Phys. Polon. B}
  {\bf 50} (2019) 1983--1992}.

\bibitem{Plenter:2020lop}
J.~Plenter and G.~Rodrigo, \emph{{Asymptotic expansions through the loop-tree duality}},  
  \href{https://doi.org/10.1140/epjc/s10052-021-09094-9}{\emph{Eur.Phys.J.C}{\bf 81} (2021) 4, 320},
  [\href{http://arxiv.org/abs/2005.02119}{{\tt 2005.02119}}].

\bibitem{Driencourt-Mangin:2019aix}
F.~Driencourt-Mangin, G.~Rodrigo, G.~F.~R. Sborlini and W.~J. Torres~Bobadilla,
  \emph{{Universal four-dimensional representation of $H \to \gamma \gamma$ at
  two loops through the Loop-Tree Duality}},
  \href{http://dx.doi.org/10.1007/JHEP02(2019)143}{\emph{JHEP} {\bf 02} (2019)
  143}, [\href{http://arxiv.org/abs/1901.09853}{{\tt 1901.09853}}].

\bibitem{Prisco:2020kyb}
R.~M.~Prisco and F.~Tramontano,
\emph{{Dual subtractions}},
\href{https://doi.org/10.1007/JHEP06(2021)089}
{\emph{JHEP} {\bf 06} (2021) 089},
[\href{http://arxiv.org/abs/2012.05012}{{\tt 2012.05012}}].

\bibitem{Hernandez-Pinto:2015ysa}
R.~J. Hernandez-Pinto, G.~F.~R. Sborlini and G.~Rodrigo, \emph{{Towards gauge
  theories in four dimensions}},
  \href{http://dx.doi.org/10.1007/JHEP02(2016)044}{\emph{JHEP} {\bf 02} (2016)
  044}, [\href{http://arxiv.org/abs/1506.04617}{{\tt 1506.04617}}].
 
\bibitem{Sborlini:2016gbr}
G.~F.~R. Sborlini, F.~Driencourt-Mangin, R.~Hernandez-Pinto and G.~Rodrigo,
  \emph{{Four-dimensional unsubtraction from the loop-tree duality}},
  \href{http://dx.doi.org/10.1007/JHEP08(2016)160}{\emph{JHEP} {\bf 08} (2016)
  160}, [\href{http://arxiv.org/abs/1604.06699}{{\tt 1604.06699}}].

\bibitem{Sborlini:2016hat}
G.~F.~R. Sborlini, F.~Driencourt-Mangin and G.~Rodrigo, \emph{{Four-dimensional
  unsubtraction with massive particles}},
  \href{http://dx.doi.org/10.1007/JHEP10(2016)162}{\emph{JHEP} {\bf 10} (2016)
  162}, [\href{http://arxiv.org/abs/1608.01584}{{\tt 1608.01584}}].

\bibitem{Driencourt-Mangin:2019sfl}
F.~Driencourt-Mangin, \emph{{Four-dimensional representation of scattering
  amplitudes and physical observables through the application of the Loop-Tree
  Duality theorem}}.
\newblock PhD thesis, U. Valencia (main), 2019.
\newblock \href{http://arxiv.org/abs/1907.12450}{{\tt 1907.12450}}.

\bibitem{Verdugo:2020kzh}
J.~J. Aguilera-Verdugo, F.~Driencourt-Mangin, R.~J. Hernandez~Pinto,
  J.~Plenter, S.~Ramirez-Uribe, A.~E. Renteria~Olivo et~al., \emph{{Open loop
  amplitudes and causality to all orders and powers from the loop-tree
  duality}},
  \href{http://dx.doi.org/10.1103/PhysRevLett.124.211602}{\emph{Phys. Rev.
  Lett.} {\bf 124} (2020) 211602}, [\href{http://arxiv.org/abs/2001.03564}{{\tt
  2001.03564}}].
  
  \bibitem{Aguilera-Verdugo:2020fsn}
J.~Jes\'us Aguilera-Verdugo, R.~J.~Hern\'andez-Pinto, G.~Rodrigo, G.~F.~R.~Sborlini and W.~J.~Torres Bobadilla, 
\emph{{Mathematical properties of nested residues and their application to multi-loop scattering amplitudes}},
\href{http://dx.doi.org/10.1007/JHEP02(2021)112}{\emph{JHEP} {\bf 02} (2021) 112}, 
 [\href{http://arxiv.org/abs/2010.12971}{{\tt 2010.12971}}].

  \bibitem{snowmass2020}
J. Aguilera-Verdugo, R. J. Hern\'andez-Pinto, S. Ram\'{\i}rez-Uribe, G. Rodrigo, G. F. R. Sborlini and W. J. Torres Bobadilla,
\emph{{Manifestly Causal Scattering Amplitudes in Snowmass 2021 - Letter of Intention}},
August 2020.
  
\bibitem{Aguilera-Verdugo:2020kzc}  
J.~J.~Aguilera-Verdugo, R.~J.~Hernandez-Pinto, G.~Rodrigo, G.~F.~R.~Sborlini and W.~J.~Torres Bobadilla,
\emph{{Causal representation of multi-loop Feynman integrands within the loop-tree duality}}.
\href{http://dx.doi.org/10.1007/JHEP01(2021)069}
{\emph{JHEP}{\bf 01} (2021)
069}, [\href{http://arxiv.org/abs/2006.11217}{{\tt 2006.11217}}].

\bibitem{Sborlini:2021owe}
G.~F.~R.~Sborlini,
\emph{{Geometrical approach to causality in multiloop amplitudes}}.
\href{https://doi.org/10.1103/PhysRevD.104.036014}
{\emph{PhysRevD}{\bf 104} (2021)
036014}, [\href{http://arxiv.org/abs/2102.05062}{{\tt 2102.05062}}].

\bibitem{TorresBobadilla:2021ivx}
W.~J.~Torres Bobadilla,
\emph{{Loop-tree duality from vertices and edges}}.
\href{https://doi.org/10.1007/JHEP04(2021)183}
{\emph{JHEP}{\bf 04} (2021)
183}, [\href{http://arxiv.org/abs/2102.05048}{{\tt 2102.05048}}].

\bibitem{Bobadilla:2021pvr}
W.~J.~T.~Bobadilla,
\emph{{Lotty \textendash{} The loop-tree duality automation}},
\href{https://doi.org/10.1140/epjc/s10052-021-09235-0}
{\emph{Eur. Phys. J. C}{\bf 81} (2021)
514}, 
[\href{http://arxiv.org/abs/2103.09237}{{\tt 2103.09237}}].

\bibitem{Ramirez-Uribe:2020hes}
S.~Ram\'\i{}rez-Uribe, R.~J.~Hern\'andez-Pinto, G.~Rodrigo, G.~F.~R.~Sborlini and W.~J.~Torres Bobadilla,
\emph{{Universal opening of four-loop scattering amplitudes to trees}}.
\href{http://dx.doi.org/10.1007/JHEP04(2021)129}
{\emph{JHEP}{\bf 04} (2021) 129},
[\href{http://arxiv.org/abs/2006.13818}{{\tt 2006.13818}}].

  \bibitem{Ramirez-Uribe:2021ubp}
S.~Ram\'\i{}rez-Uribe, A.~E.~Renter\'\i{}a-Olivo, G.~Rodrigo, G.~F.~R.~Sborlini and L.~Vale Silva,
\emph{{Quantum algorithm for Feynman loop integrals}}.
\href{http://arxiv.org/abs/2105.08703}{{\tt 2105.08703}}.

\bibitem{Bollini:1972ui}
C.~G. Bollini and J.~J. Giambiagi, \emph{{Dimensional Renormalization: The
  Number of Dimensions as a Regularizing Parameter}},
  \href{http://dx.doi.org/10.1007/BF02895558}{\emph{Nuovo Cim.} {\bf B12}
  (1972) 20--26}.

\bibitem{tHooft:1972tcz}
G.~'t~Hooft and M.~J.~G. Veltman, \emph{{Regularization and Renormalization of
  Gauge Fields}},
  \href{http://dx.doi.org/10.1016/0550-3213(72)90279-9}{\emph{Nucl. Phys.} {\bf
  B44} (1972) 189--213}.
\bibitem{Boyer:1996zf}
M.~Boyer, G.~Brassard, P.~Hoyer and A.~Tapp,
\emph{{Tight bounds on quantum searching}},
\href{https://doi.org/10.1002/(SICI)1521-3978(199806)46:4/5<493::AID-PROP493>3.0.CO;2-P}
{\emph{Fortsch. Phys.} {\bf 46} (1998) 493-506},
[\href{http://arxiv.org/abs/quant-ph/9605034}{{\tt
  quant-ph/9605034}}].

\bibitem{Grover:1997fa}
L.~K.~Grover, 
\emph{{Quantum mechanics helps in searching for a needle in a haystack}},
  \href{http://dx.doi.org/10.1103/PhysRevLett.79.325}{\emph{Phys. Rev.
  Lett.} {\bf 79} (1997) 325-328}, [\href{http://arxiv.org/abs/quant-ph/9706033}{{\tt
  quant-ph/9706033}}].

\bibitem{alonso_raul_2021_5561050}
R.~Alonso, A.~Arias, P.~Coca, F.~D\'iez, A.~Garc\'ia and L.~Meijueiro, \emph{Qute: Quantum computing simulation platform},  Oct., 2021.
\newblock 10.5281/zenodo.5561050.


\end{thebibliography}
\end{document}